\documentclass[aps,prl,nobibnotes,twocolumn,superscriptaddress,bibliography]{revtex4-1}
\usepackage{amsfonts}
\usepackage{mathrsfs}
\usepackage{amsmath}
\usepackage{color}
\usepackage{natbib}
\usepackage{graphicx}
\usepackage{bm}
\usepackage{amssymb}
\usepackage{xspace}
\usepackage{epstopdf}
\usepackage{dcolumn}
\usepackage{longtable}
\usepackage{array}
\usepackage{makecell}
\usepackage{tabulary}
\usepackage{multirow}
\usepackage{appendix}
\usepackage{mathtools}
\usepackage{pifont}
\usepackage{longtable}
\usepackage{float}
\newcolumntype{C}[1]{>{\centering\arraybackslash}p{#1}}
\newcolumntype{L}[1]{>{\flushleft\arraybackslash}p{#1}}

\usepackage{multirow}
\usepackage{epsfig}
\usepackage[normalem]{ulem}
\usepackage{amsmath}
\usepackage{braket}
\usepackage{bbold}
\usepackage{graphicx}

\makeatletter
\usepackage[colorlinks=true, letterpaper=true, pdfstartview=FitV, linkcolor=blue, citecolor=blue, urlcolor=blue]{hyperref}
\makeatother

\begin{document}
\title{Interstitial-Electron Altermagnetism in Two Dimensions}

\author{Xia Cheng}\thanks{These authors contributed equally to this work.}
\address{School of Physical Science and Technology, Southwest University, Chongqing, 400715 China.}
\address{Institute for Superconducting and Electronic Materials, Faculty of Engineering and Information Sciences, University of Wollongong, Wollongong 2500, Australia.}
\author{Yang Wu}\thanks{These authors contributed equally to this work.}
\address{Bremen Center for Computational Materials Science, University of Bremen, 28359 Bremen, Germany.}
\author{Zhenzhou Guo}
\address{Institute for Superconducting and Electronic Materials, Faculty of Engineering and Information Sciences, University of Wollongong, Wollongong 2500, Australia.}
\author{Tie Yang}
\address{School of Physical Science and Technology, Southwest University, Chongqing, 400715 China.}
\author{Weizhen Meng}%\thanks{Corresponding authors}\email{mengweizhen@hebtu.edu.cn}
\address{College of Physics, Hebei Key Laboratory of Photophysics Research and Application, Hebei Normal University, Shijiazhuang 050024, China.}
\author{Zhenxiang Cheng}
\address{Institute for Superconducting and Electronic Materials, Faculty of Engineering and Information Sciences, University of Wollongong, Wollongong 2500, Australia.}
\author{Zhi-Ming Yu}%\thanks{Corresponding authors}\email{zhiming$\_$yu@bit.edu.cn}
\address{School of Physics, Beijing Institute of Technology, Beijing, 100081 China.}
\author{Xiaotian Wang}%\thanks{Corresponding authors}\email{xiaotianw@uow.edu.au}
\address{Institute for Superconducting and Electronic Materials, Faculty of Engineering and Information Sciences, University of Wollongong, Wollongong 2500, Australia.}

\begin{abstract}
Altermagnetism has so far been associated with compensated magnetic moments carried by atoms. Here we introduce Stoner instability induced interstitial-electron altermagnetism, a distinct mechanism in which altermagnetic order is carried instead by interstitial anionic electrons in electrides. We show that, owing to the quasi-nucleus-free nature of interstitial electrons, the Stoner instability in electrides hosting two interstitial electrons can naturally stabilize an altermagnetic state rather than the conventional ferromagnetic one. This mechanism leads to a practical design principle for two-dimensional materials, from which we identify monolayers Zr$_{2}$N and Ti$_2$N as representative candidates. The strong sensitivity of interstitial electrons to cavity size enables efficient strain control of the altermagnetic order and a pronounced piezo-altermagnetic effect. Moreover, we investigate the evolution of the magnetism in Zr$_{2}$N under ultrafast laser excitation, which exhibits dynamics distinct from those in all previously reported magnetic materials where magnetism is carried by real atoms. Our work not only offers a novel pathway to realize altermagnetism but also reveals an efficient non-magnetic route for its control.
\end{abstract}
\maketitle

%%%%%%% Main text %%%%%%%%%%%%%%%%%%%%%
\textit{\textcolor{blue}{Introduction---}}Electrides represent a class of unconventional electron-rich materials characterized by excess electrons that are highly localized within structural cavities and act as interstitial anionic electrons (IAEs)~\cite{dye1993anionic,dye2003electrons,matsuishi2003high}. In electrides, the dimensionality of these IAEs---governed by the spatial confinement of the interstitial sites---can range from zero to three, corresponding to $n$D  electrides with $n$ = 0, 1, 2, and 3~\cite{hosono2021advances,meng2025magnetic}. Notably, the quasi-nucleus-free nature of these IAEs endows the host materials with remarkable properties, including low work functions~~\cite{toda2007work,lee2013dicalcium,hirayama2018electrides,wan2024bacu}, diverse topological states~\cite{hirayama2018electrides,park2018first,huang2018topological,liu2020ferromagnetic,meng20251d}, unique magnetic behavior~\cite{lee2020ferromagnetic,zhang2023magnetic,hwang2025permanent}, and high-pressure superconductivity~\cite{miyakawa2007superconductivity,zhao2019predicted,liu2021proposed}. These properties render  electrides promising candidates for applications in light-emitting diodes, energy storage and conversion systems, and electron emission devices~\cite{kitano2012ammonia,hosono2021advances,gong2022unique,yang2024vacancy}.

Meanwhile, the recent proposal of altermagnetism has attracted significant attention~\cite{add1,add2,bai2024altermagnetism,song2025altermagnets}. Although altermagnetic (AM) materials possess no net magnetization, they exhibit band spin splitting reminiscent of ferromagnets: specifically, altermagnets can host alternating spin-split electronic bands in the absence of spin-orbit coupling (SOC). A variety of intriguing phenomena associated with AM  systems have since been reported~\cite{vsmejkal2020crystal,Mazin2021,add27,feng2022anomalous,H-Bai2022,Karube2022,H-Bai2023,add10,add15,XD-Zhou2024,amin2024nanoscale,add19,add20,ZY-Zhou2025,zhou2025manipulation}.
Moreover, many  theoretical proposals, such as stacking, twisting, or external fields~\cite{liu2024twisted,zhu2025design,duan2025antiferroelectric,sun2025proposing,che2025engineering,che2025bilayer,liu2025realizing,zhu2025floquet,huang2025light,che2026symmetry}, have been put forward to realize 2D AM materials. Notably, in all previous proposals, the AM  behavior originates from real magnetic atoms. In contrast, altermagnetism induced by IAEs has remained  unexplored, largely because a general mechanism to control the magnetism of IAEs is still lacking. Furthermore, the intriguing properties of electrides generally are often attributed to the isolated nature of IAEs. In realistic materials, however, IAEs cannot be completely decoupled from the surrounding lattice, and such IAE--lattice coupling may open up new possibilities for efficiently manipulating IAEs-induced magnetism; yet this aspect has not been explored.

\begin{figure}[t]
\centering
\includegraphics[width=1\linewidth]{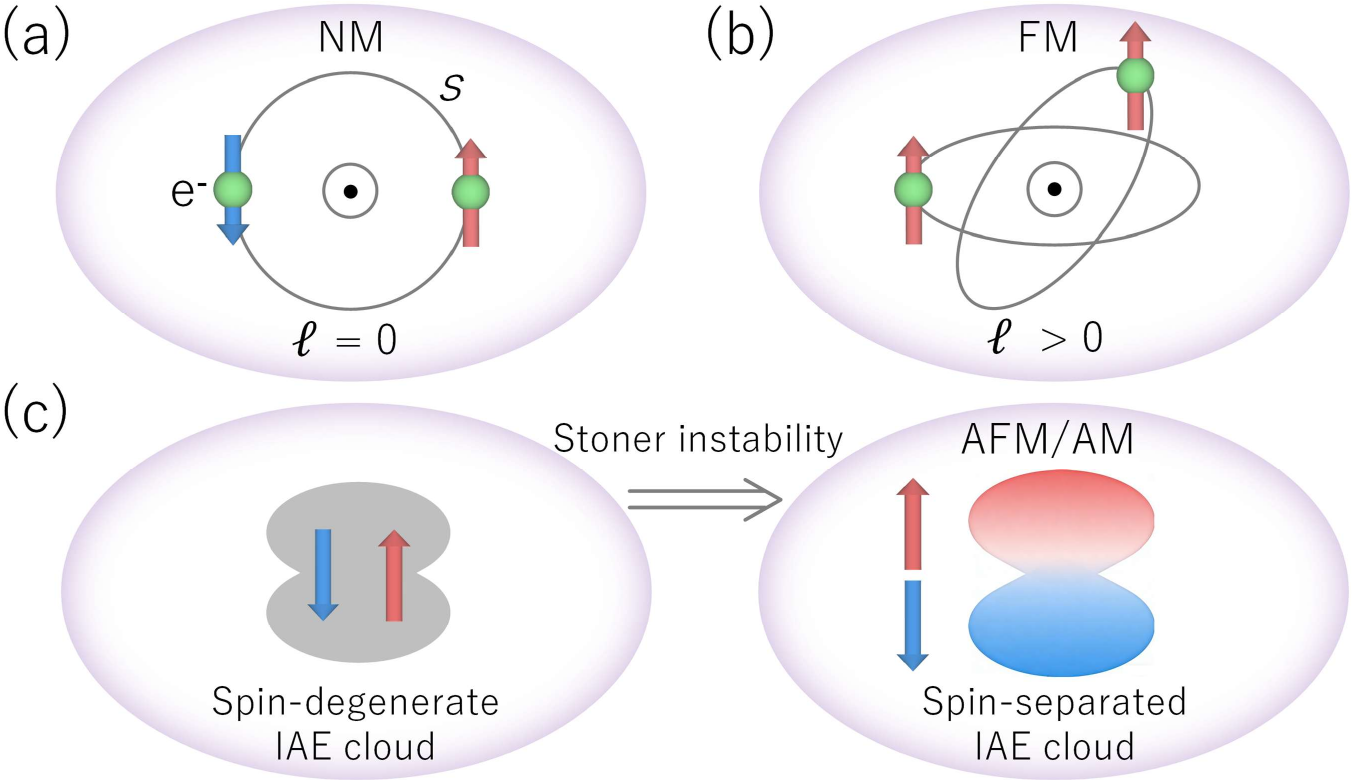}
\caption{(a-b) An ion with two electrons in the outermost shell. %The two electrons are tightly  bound to the  real ions and localized in the same position.
Since the electrons  are tightly  bound to the  real ions, the electron clouds for the two electrons are confined in the same position. They have to choose suitable orthogonal orbits to lower the ground state energy.
In (a) and (b), the  outermost shells  are the  orbits  with $\ell = 0$ and $\ell >0$, respectively.
Accordingly to Hund's rule, the orbit with $\ell = 0$ ($\ell > 0$)  is  fully (partially) occupied, leading to  a nonmagnetic  (magnetic) state in (a) [(b)].
%(a) An atom possessing only an $\emph{s}$ orbital exhibits nonmagnetic behavior when that orbital is fully occupied ($\ell = 0$, where $\ell$ denotes the orbital angular momentum quantum number). (b) An atom with high-energy level orbitals (e.g., $\emph{p}$, $\emph{d}$, or $\emph{f}$) exhibits ferromagnetic behavior when those orbitals are partially occupied ($\ell > 0$).
(c) In  electrides, the IAEs are not tightly  bound to certain ions, but actually  delocalize throughout the unit cell, only with more weight distributed  within certain cavities. Without the binding constraint imposed by the ions, the IAEs can adopt a fundamentally different way to lower the ground-state energy: under  Stoner instability,  the spin-degenerate IAE clouds can directly move in opposite directions.
Such spatial separation of the spin-polarized IAE clouds  lead to antiferromagnetism or altermagnetism.
\label{fig1}}
\end{figure}

In this Letter, we identify a previously unreported type of altermagnetism driven by the Stoner instability and hosted by IAEs rather than by atoms. Our proposal is to explore metallic electrides in 2D containing two excess electrons. In such systems, IAEs generally can generate pronounced density peaks in the density of states (DOS) at the Fermi level, thereby promoting a Stoner instability. Remarkably, owing to their quasi-nucleus-free character, the two IAEs in these electrides can lower the ground-state energy in an unconventional manner--by separating in real space and adopting antiparallel spin alignment, as illustrated in Fig.~\ref{fig1}(c). Consequently, conventional Stoner ferromagnetism is suppressed and replaced by a Stoner-driven altermagnetism, provided that the crystal structure lacks spatial inversion symmetry. We demonstrate our idea using the monolayer electrides Zr$_2$N and Ti$_2$N, which serve as prototypical realizations of our design principle. Particularly, monolayer Zr$_2$N (Ti$_2$N) is metallic and exhibits sizable AM spin splitting. The magnetic moment associated with the IAEs reaches approximately $\sim 0.109~\mu_B$ ($\sim 0.260~\mu_B$) for monolayer Zr$_2$N (Ti$_2$N). Furthermore, we identify pronounced IAE--lattice coupling in monolayer Zr$_2$N, where the AM order parameter is sensitive to the lattice constants. A magnetic phase transition from AM to nonmagnetic state occurs under a biaxial strain  of -3$\%$, showing a piezo-altermagnetic effect. The unique evolution of magnetism in Zr$_2$N under ultrafast laser excitation has also been discussed, which has not been observed in previous studies of magnetic materials where magnetism is carried by real atoms.

\textit{\textcolor{blue}{AM electrides---}}A key characteristic of metallic  electrides is that IAEs typically generate a pronounced DOS peak at the Fermi level, thereby promoting a Stoner instability. Interestingly, the magnetism arising from Stoner instability in IAEs can be fundamentally different from that of electrons tightly bound to real ions. Consider an ion with two electrons in the outermost shell. Two distinct cases arise. If both electrons occupy an $s$ orbital, the $s$ level is fully filled, resulting in a nonmagnetic state, as the two electrons with opposite spins are confined to the same position [see Fig.~\ref{fig1}(a)]. If outermost shell is not $s$-orbit, we generally obtain a high-spin state with ferromagnetism following Hund's rule [see Fig.~\ref{fig1}(b)].

In sharp contrast, the two IAEs in 2D systems can lower the ground-state energy through a distinct mechanism. When two electrons occupy the same position, the electron-electron Coulomb interaction becomes significant. The most natural way to reduce the Coulomb interaction is for the electrons to separate in real space--a process that is generally impossible for electrons around real ions, as they are tightly bound to the ions. However, this scenario can emerge in electrides due to the weaker constraint imposed by the ions. Fig.~\ref{fig1}(c) shows an electride with two IAEs. The IAE is, in fact, an electron cloud that has certain possibility to appear in any position of the cavity of the lattice. In the nonmagnetic configuration, the IAE cloud remains spin-degenerate [see Fig.~\ref{fig1}(c)], which typically corresponds to a higher ground-state energy. The system can lower its energy by shifting the two IAE clouds in opposite directions. Moreover, the Pauli exclusion principle favors antiparallel spin alignment in this separated configuration, further stabilizing the state. As a consequence, the two IAEs in electrides may adopt a Stoner-instability-driven AM or antiferromagnetic (AFM) configuration, depending on the crystalline structure of the system.
\begin{figure}[t]
\centering
\includegraphics[width=1\linewidth]{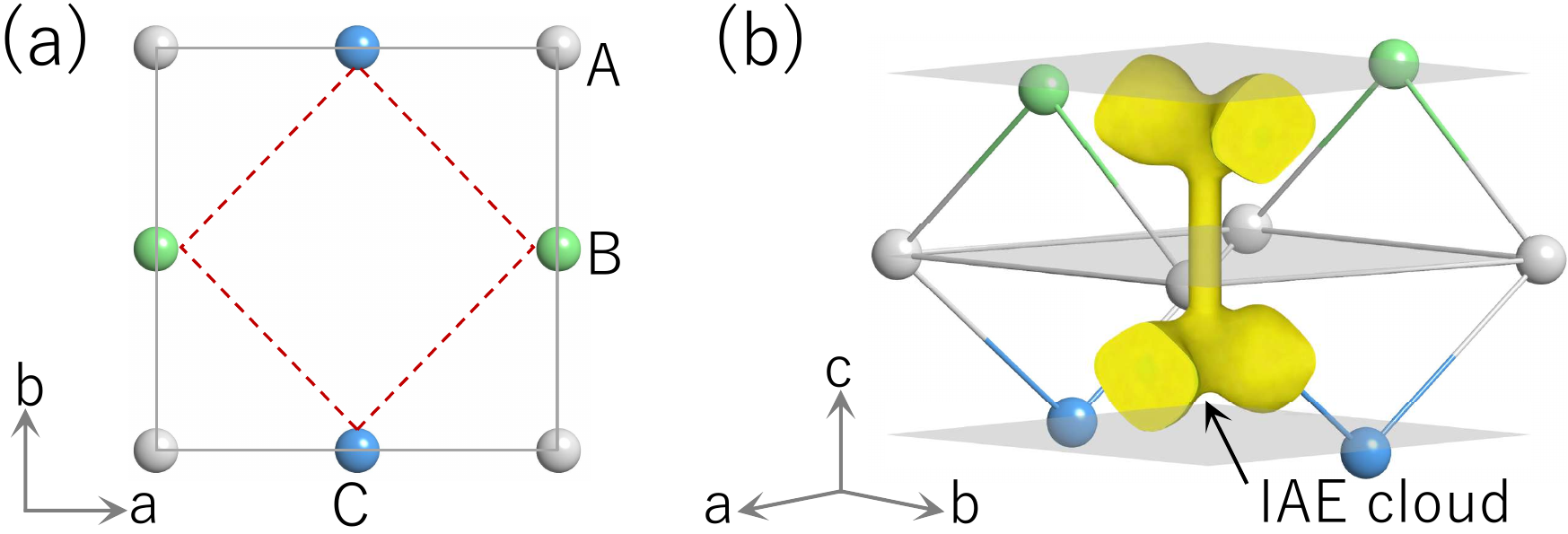}
\caption{(a) Top  and (b) side view of a modified monolayer Lieb lattice composed of three atomic species. The red region in (a) denotes a cavity in the  lattice.  The yellow region in (b) represents a possible IAE cloud in the cavities of the  lattice.
\label{fig2}}
\end{figure}

\textit{\textcolor{blue}{Design strategy---}}We begin with a modified 2D Lieb lattice [see Fig.~\ref{fig2}(a)], whose geometry provides a natural platform for hosting electrides with two IAEs. A conventional 2D Lieb lattice contains three atoms per unit cell and preserves spatial inversion symmetry ${\cal{P}}$, as all atoms lie within the same plane. To break ${\cal{P}}$ and enable interstitial-electron AM, we displace atoms B and C in opposite out-of-plane directions while keeping atom A in the original middle plane, as illustrated in Fig.~\ref{fig2}(b). This structural modification removes ${\cal{P}}$ and simultaneously generates a sizable cavity in each unit cell [see Fig.~\ref{fig2}], which is essential for stabilizing IAEs. To accommodate two IAEs per unit cell, a valence imbalance of two electrons between cations and anions is required. For structural stability and charge neutrality, atoms B and C are chosen as identical cations, supplying two excess electrons per unit cell. When the two IAEs simultaneously adopt an  AFM configuration, the ${\cal{P}}$-broken lattice symmetry allows a nonrelativistic spin splitting in momentum space, thereby realizing an interstitial-electron AM electride.

\begin{figure}[t]
\centering
\includegraphics[width=0.96\linewidth]{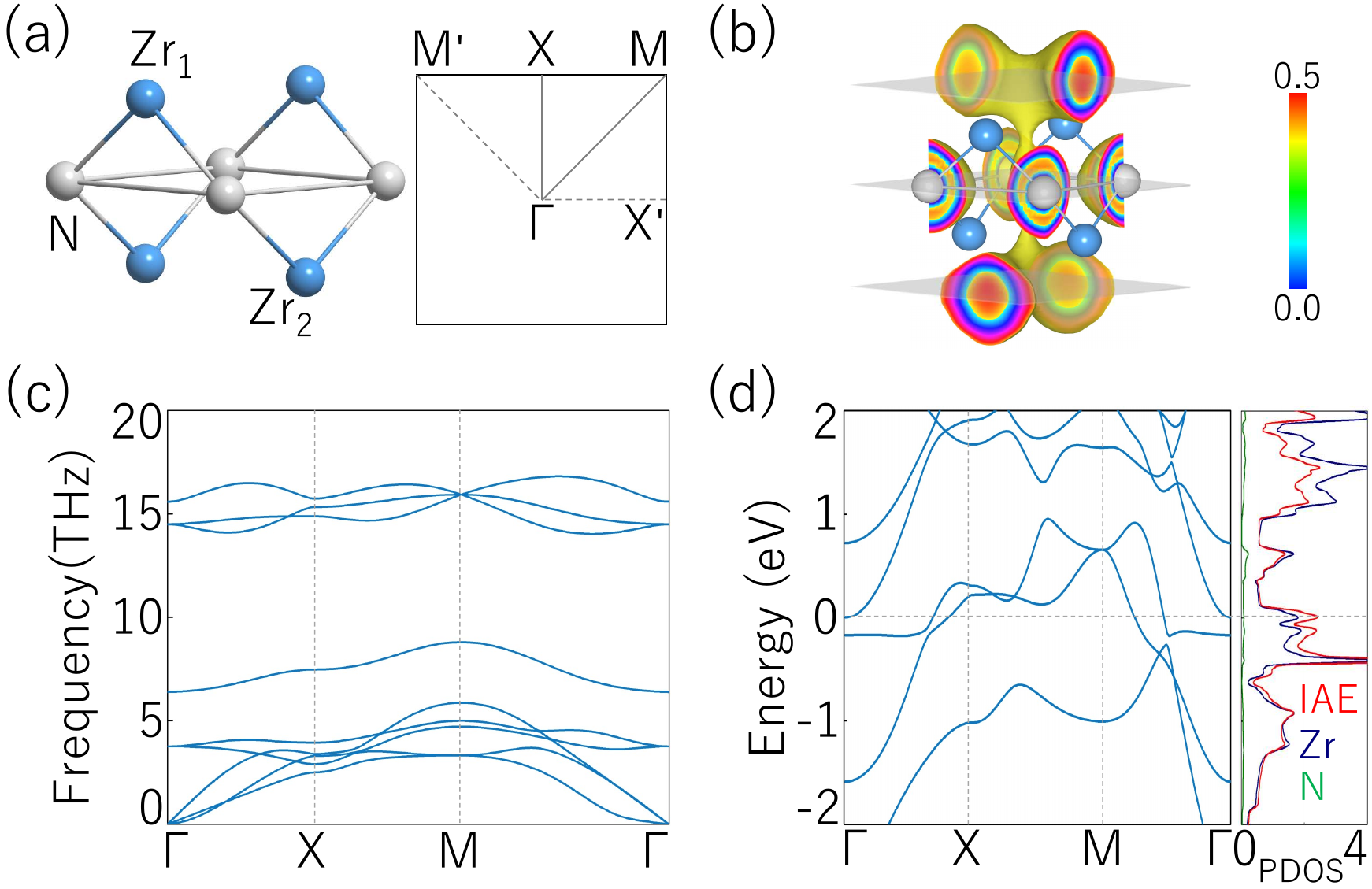}
\caption{(a) The crystal structure and corresponding Brillouin zone of Zr$_{2}$N. (b-d) The properties of Zr$_{2}$N in the nonmagnetic state. (b) is the side view of the electron localization function (ELF), (c) plots the phonon spectrum, and  (d) shows the electronic band structure and partial density of states (PDOS).
\label{fig3}}
\end{figure}

\textit{\textcolor{blue}{Material realization---}}Based on the proposed design strategy, we identify monolayers Zr$_{2}$N and Ti$_2$N as ideal candidates. Since the two systems exhibit nearly identical physical properties, we focus on Zr$_{2}$N in the main text, while the corresponding results for Ti$_2$N are provided in the Supplementary Material (SM) {[see Fig. S1 of the SM~\cite{SM}]}. In Fig.~\ref{fig3}(a), we present the crystal structure of the monolayer Zr$_{2}$N, which adopts a modified Lieb lattice configuration. Each unit cell consists of two Zr atoms located in the top and bottom layers, respectively, and one N atom positioned in the middle layer. The crystalline structure of monolayer Zr$_{2}$N belongs to space group No. 115 (${P{\bar 4}m2}$), which lacks ${\cal{P}}$. A qualitative valence-state analysis, based on the common oxidation states of Zr ($+2$, $+3$, or $+4$) and the typical $-3$ valence of N in metal nitrides, suggests that monolayer Zr$_2$N is intrinsically electron-rich. To quantify this, we performed Bader charge analysis using the BadELF method~\cite{weaver2023counting}. The results indicate that Zr atoms adopt a mixed valence state between +2 and +3. While part of the electrons donated by Zr atoms is transferred to N atom, the remaining electrons localize at interstitial regions, forming IAEs. The resulting approximate charge distribution is +1.75 $|$e$^{-}$$|$  on each Zr atom, -1.50 $|$e$^{-}$$|$ on the N atom, and -2.00 $|$e$^{-}$$|$ associated with the IAEs.

The electron localization function (ELF) of nonmagnetic Zr$_2$N is shown in Fig.~\ref{fig3}(b), where two localized IAE clouds are clearly visible, primarily distributed within the lattice cavities between the top and bottom Zr layers. These IAE clouds remain spin degenerate due to the nonmagnetic feature of this state. Upon the onset of Stoner instability, the system will lower its energy by lifting the spin degeneracy, resulting in spatial separation of the two IAE clouds with antiparallel spin alignment, as shown below.

\begin{figure}[t]
\centering
\includegraphics[width=1\linewidth]{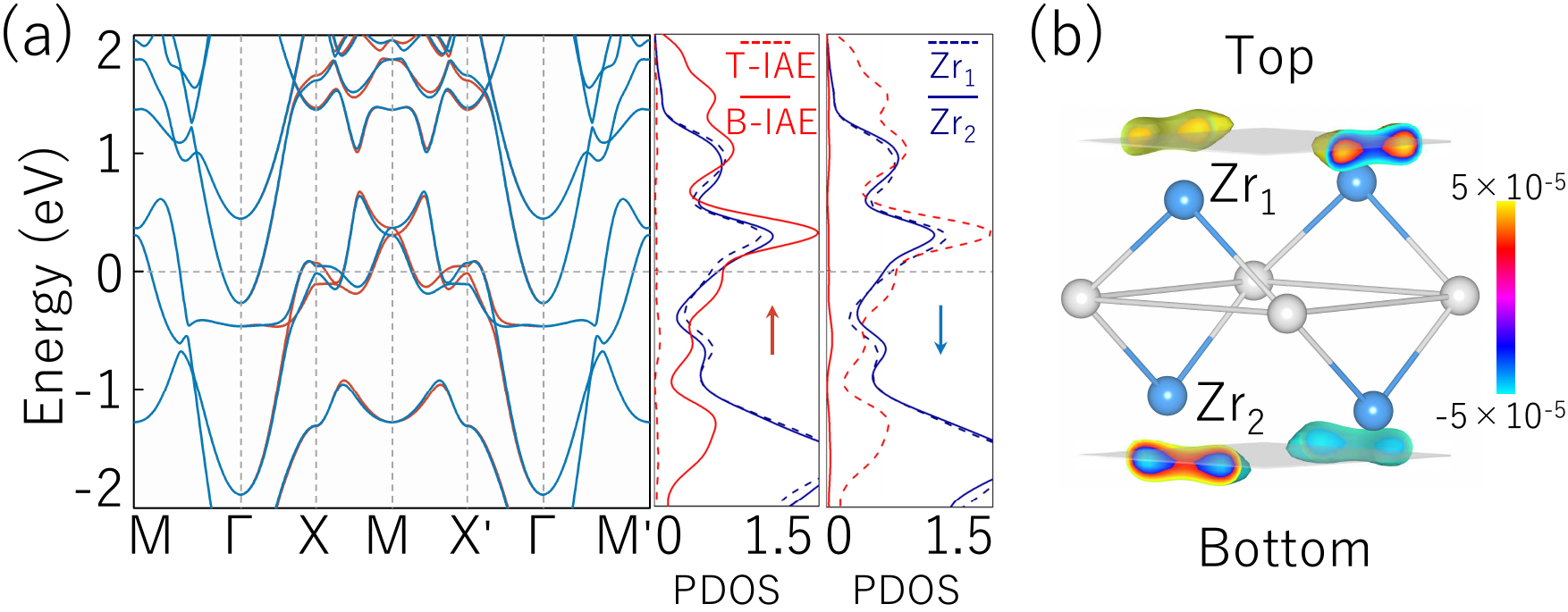}
\caption{(a) Electronic band structure and PDOS of Zr$_{2}$N in the AM state. (b) Spin density of Zr$_{2}$N in the AM order.
\label{fig4}}
\end{figure}

We then examine the dynamical stability of monolayer Zr$_2$N. As shown in Fig.~\ref{fig3}(c), Zr$_2$N is dynamically stable in the nonmagnetic state, as evidenced by the absence of imaginary phonon modes throughout the Brillouin zone. Nevertheless, dynamical stability does not necessarily imply that the nonmagnetic phase represents the true ground state. The electronic band structure of the nonmagnetic monolayer is shown in Fig.~\ref{fig3}(d), revealing a clear metallic character. An important feature of the nonmagnetic state is a pronounced DOS peak [$D(E_F)$] at the Fermi level, a typical feature of metallic electrides. Within the Stoner framework, such an enhanced $D(E_F)$ promotes magnetic instability, driving the system toward spin polarization to reduce the density of states at $E_F$ and lower the total energy. We therefore examine three possible magnetic configurations of the two IAEs: nonmagnetic, ferromagnetic (FM), and AM. Our calculations show that, for the monolayer Lieb-lattice Zr$_2$N, the AM state possesses the lowest total energy, accompanied by a suppressed DOS at the Fermi level, as shown in Fig.~\ref{fig4}(a) and Fig. S2 of the SM~\cite{SM}. In particular, detailed analysis shows that  the two IAE clouds are no longer coincident in real space. Instead, the spin-up (spin-down) IAE cloud contributes predominantly to the cavity in the top (bottom) layer, as illustrated in Fig.~\ref{fig4}(b). This AM order and spatial separation of the IAE clouds is fully consistent with our previous qualitative analysis. We further confirm that the AM phase of Zr$_2$N remains dynamically stable (see Fig. S3 of the SM~\cite{SM}).

The spin space group of AM Zr$_2$N is \( P^{\bar{1}}\bar{4}^{1}m^{\bar{1}}2^{\infty m}1 \) with identifier 115.25.1.1.L~\cite{PhysRevX.14.031038}, indicating that the two spin-opposite IAEs are related by the combined symmetry [$\mathcal{C}_{2}\parallel\mathcal{S}_{4}$] symmetry, rather than [$\mathcal{C}_{2}\parallel\mathcal{P}$] symmetry or a time reversal symmetry ${\cal{T}}$ followed by a fractional translation. The AM nature of monolayer Zr$_2$N is further evidenced by its band structure, where characteristic AM spin splitting is clearly observed along the $\Gamma$--X--M--X$'$--$\Gamma$ path, as shown in Fig.~\ref{fig4}(a), while the bands remain spin-degenerate along the $\Gamma$--M and $\Gamma$--M$'$ directions. We further confirm the electride character and the robustness of the AM metallic state using HSE hybrid functional calculations (see Fig. S4 of the SM~\cite{SM}).

One can understand the nonrelativistic spin splitting of the IAE-induced altermagnet via a magnetic multipole
\begin{equation}
\mathbf{O}_{\mu\nu} = \int d^3 r \,r_{\mu} r_{\nu}\, m(\mathbf{r}),
\end{equation}
where $m(\mathbf{r})$ denotes the microscopic magnetization density, and $\mu,\nu \in \{x,y,z\}$ label the Cartesian components of the position vector $\mathbf{r}$. In previously reported altermagnets, $m(\mathbf{r})$ is mainly distributed around magnetic atoms~\cite{bhowal2024ferroically,
mcclarty2024landau,liu2025multipolar}. In contrast, in the present system  $m(\mathbf{r})$ is dominated by spin-polarized IAEs residing in the crystal cavities, as shown in Fig.~\ref{fig4}(b). For clarity, the spin density of these IAEs exhibits a spatially anisotropic distribution between the top and bottom interstitial sites, giving rise to a finite magnetic multipole while maintaining a vanishing net magnetization. The band spin splitting exhibits a form consistent with the magnetic multipole. As shown in Fig.~\ref{fig4}(a), the electronic structure displays the characteristic nonrelativistic spin splitting
\begin{equation}
\Delta E(\mathbf{k}) \propto k_x^2 - k_y^2.
\end{equation}
 These results indicate that the AM oder in this system is governed primarily by IAEs rather than real atoms, revealing an IAEs--driven mechanism of AM.

The AM nature of Zr$_2$N can also be inferred from the projected density of states (PDOS). The PDOS analysis shows that the electronic states near the Fermi level in both spin channels are primarily derived from the IAEs and the Zr $d$ orbitals [see Fig.~\ref{fig4}(a)]. Notably, the DOS of the top and bottom Zr atoms is nearly identical across the entire energy range for both spin channels. In contrast, the IAEs in the top and bottom layers exhibit pronounced spin-dependent differences. This indicates that the altermagnetism in Zr$_2$N originates predominantly from the IAEs. Furthermore, the magnetic moments associated with the two IAEs and the two Zr atoms are explicitly evaluated. We find that the AM order is predominantly carried by the IAEs: the IAEs in the top and bottom layers possess magnetic moments of approximately  $\sim\mathrm{0.109}~\mu_B$ and $\sim-\mathrm{0.109}~\mu_B$, respectively, resulting in a vanishing net magnetic moment. In contrast, the magnetic moments on the top and bottom Zr atoms are much smaller ($\sim\pm0.027~\mu$$_{B}$).

\begin{figure}
\centering
\includegraphics[width=0.97\linewidth]{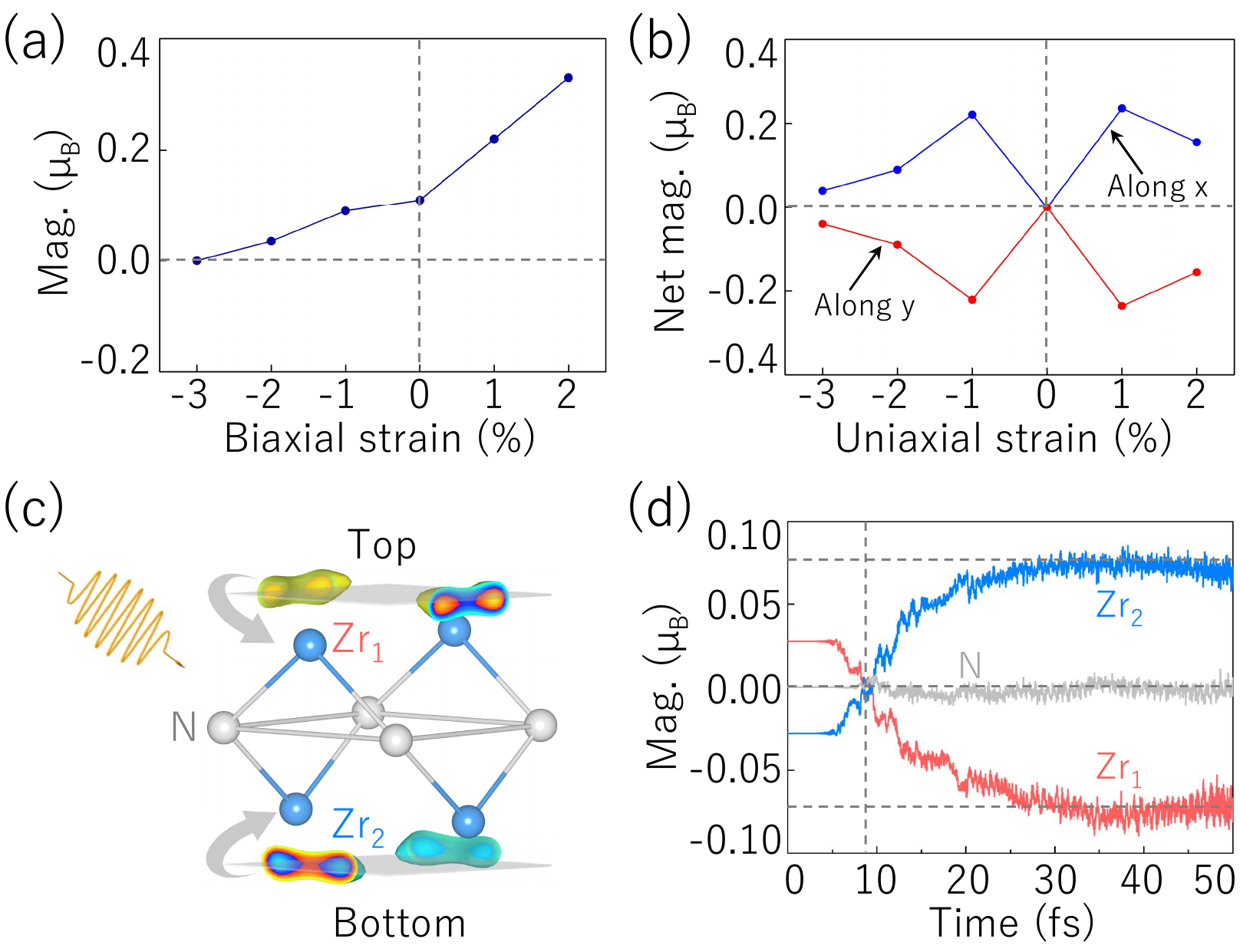}
\caption{(a) Evolution of the absolute magnetic moment of each IAE in the AM Zr$_{2}$N under biaxial strain. (b) Evolution of the net magnetic moment in Zr$_{2}$N under uniaxial strain along the x and y directions. (c) Schematic diagram of ultrafast laser-induced spin transfer and reversal in Zr$_{2}$N. (d) Time-dependent dynamics of the magnetic moments for Zr$_{1}$ atom (top layer, red), Zr$_{2}$ atom (bottom layer, blue) and N atom (middle layer, grey).
\label{fig5}}
\end{figure}

\textit{\textcolor{blue}{IAE--lattice coupling---}}Since the IAEs are highly sensitive to the size of the structural cavities, strain provides an effective means to tune their properties, thereby establishing a strong coupling between the IAEs and the lattice. This coupling becomes particularly intriguing when the IAEs carry finite magnetic moments, as it may give rise to an unconventional piezo-magnetic effect. This behavior can be directly demonstrated under biaxial strain. As shown in Fig.~\ref{fig5}(a), the monolayer Zr$_{2}$N keeps AM order when the biaxial strain is less than $-3\%$. However, the AM order parameter--defined as the absolute magnetic moment of each IAE--gradually decreases with increasing compressive strain. Particularly, for a critical value of $\sim -3\%$, the monolayer Zr$_{2}$N undergoes a magnetic  phase transition from AM state to nonmagnetic state. These results reveal a pronounced and tunable piezo-altermagnetic effect in monolayer Zr$_2$N.

In addition, applying uniaxial strain breaks the [$\mathcal{C}_{2}\parallel\mathcal{S}_{4}$] symmetry that connects the two spin-opposite IAEs. Once this symmetry constraint is lifted, the two IAEs are no longer symmetry-related, and the system undergoes a transition from the AM state to the FM state. Furthermore, the ferromagnetism of the strained Zr$_{2}$N can be switched by applying uniaxial strain along the $x$ and $y$ directions [see Fig.~\ref{fig5}(b)].

\textit{\textcolor{blue}{Evolution under ultrafast laser excitation---}}We further investigate the ultrafast evolution of the magnetism of  the monolayer Zr$_{2}$N under laser excitation. Using real-time time-dependent density functional theory (rt-TDDFT) simulations, we confirm the presence of magnetic IAEs and reveal a transient magnetic-moment transfer from the IAEs to the Zr atoms, as illustrated in Fig.~\ref{fig5}(c). The element-resolved spin dynamics exhibit two distinct stages [Fig.~\ref{fig5}(d)]. (i) During 5-8 fs, the initially weak-magnetic Zr atoms undergo rapid demagnetization, where the magnetic moment changes from 0.027 $\mu$$_{B}$ to nearly zero; (ii) After 8 fs, the local magnetic moment of the Zr atoms in the top (bottom) layer undergoes spin switching, followed by a steady increase in magnitude. The magnetic moments eventually stabilize at approximately $\pm$0.086 $\mu$$_{B}$, significantly exceeding the ground-state intrinsic value and approaching the magnetization magnitude of the IAEs. This enhancement originates from spin transfer from the IAEs, as evidenced by the evolution of the spin density (see Fig. S5 of the SM~\cite{SM}). Notably, such ultrafast spin reversal accompanied by magnetic-moment enhancement is absent in previously reported ultrafast demagnetization processes of conventional magnets lacking IAEs~\cite{geneaux2024spin,harris2024spin,zhou2025ultrafast}.

\textit{\textcolor{blue}{Discussion---}}We establish a previously unexplored form of altermagnetism hosted by IAEs in electrides with a  modified monolayer Lieb lattice. Owing to their quasi-nucleus-free character, Stoner instability induces real-space separation of two IAE clouds with antiparallel spin alignment, giving rise to AM spin splitting without net magnetization. We demonstrate this mechanism in monolayers Zr$_{2}$N and Ti$_{2}$N, where the AM order originates predominantly from IAEs rather than real atoms. Due to strong IAE--lattice coupling, the altermagnetism uncovered here can be efficiently tuned by strain, leading to a pronounced piezo-altermagnetic response. Finally, ultrafast excitation reveals spin dynamics in monolayer Zr$_2$N beyond those of conventional magnets lacking IAEs, highlighting the active role of IAEs in ultrafast laser control.

Experimentally, similar to the synthesized layered electride Ca$_2$N~\cite{lee2013dicalcium}, Hall measurements reveal that the charge carriers originate from IAEs, with carrier densities consistent with the stoichiometric value, indicating that IAEs in monolayer Zr$_2$N may be reliably identified through carrier-density measurements. Moreover, similar to the spin-split band structures verified in 2D Rb-doped V$_2$Te$_2$O~\cite{zhang2025crystal} and V$_2$Se$_2$O~\cite{jiang2025metallic}, angle-resolved photoemission spectroscopy (ARPES) measurements may directly resolve the predicted AM spin splitting in monolayer Zr$_2$N.

\textit{Acknowledgments---}This work was supported by the Australian Research Council Discovery Early Career Researcher Award (Grant No. DE240100627), the National Natural Science Foundation of China (Grants No. 12474040 and 12404073), the Australian Research Council Discovery Project (Grant No. DP260102992), the Science Research Project of Hebei Education Department (Grant No. BJ2025051), and the China Postdoctoral Science Foundation (Grant No. 2025M773382). We acknowledge the allocation of high-performance computing time through the UOW Partner Share scheme. We also acknowledge the computational resources provided by the National Computational Infrastructure (NCI), which were allocated under the National Computational Merit Allocation Scheme supported by the Australian Government.

\bibliography{ref}

\end{document}